\documentclass[12pt]{article}
\begin{document}
\title{The Astronomy and Science Behind the Myth}
\author{B.G. Sidharth\\
International Institute for Applicable Mathematics \& Information Sciences\\
Hyderabad (India) \& Udine (Italy)\\
B.M. Birla Science Centre, Adarsh Nagar, Hyderabad - 500 063
(India)}
\date{}
\maketitle
\begin{abstract} We show that many ancient Indian
mythological legends are really allegorical depictions of
astronomical and other physical phenomena of which the composers had
a good idea. This detail enables us to also use techniques of
astronomical chronology for dating various events.
\end{abstract}
The word myth can be traced to the Sanskrit word Mithya which
literally means untruth. Yet much of Indian mythology is a poetic
and allegorical description based on hard core astronomical and
scientific facts and events \cite{go1,go2,ck}. Glimpses of this
truth in a wider though less precise context have also been noted in
reference \cite{santilana}. Unfortunately, the fact that most of the
scholars of mythology have had a background in languages and
literature rather than science has kept the inner scientific meaning
masked. A correct interpretation requires a knowledge of science and
particularly Astronomy. Such an interpretation is very revealing and
makes many obscure notions meaningful. Let us take a few examples
which bring this to
light.\\
In the Rig Veda, which is the oldest extant Indo-European
literature, there are a number of hymns attributed to Sunashepa,
literally the tail of the dog \cite{rigveda}. From these hymns it
appears that Sunashepa was to be sacrificed to the heavens, as
promised by his father. He is bound by three pegs. As the time of
the sacrifice approaches Sunashepa implores the Gods and finally is
spared, though the head of an animal is accepted instead. A similar
myth occurs later in other traditions also. Shorn of all the
imagery, poetry and myth, this is the description of a heavenly
drama which we can witness even today: Orion the Hunter, with three
stars on his belt and the dog Sirius trailing behind, trying to
enter the zodiacal belt of the heavens. Interestingly the stars near
Orion's head are, in Sanskrit called Mrigasira or literally head of
an animal.\\
This story of Sunashepa reappears in two variants in later Hindu
epics. The first in the elaborate story of Trisanku in the Ramayana,
literally with three pegs or knots, a mythical king who tried
unsuccessfully to enter heaven, that is the zodiac, in his human
form. He is not admitted and is left hanging midway. Indeed apart
from being mere mythological representations of the drama of the
constellations, all this also yields valuable astronomical
chronology that enables us to date these epics and events, as argued
by the author \cite{ck}.\\
Yet another appearance of the same myth is in the later epic, the
Mahabharata where the just king Yudhishtra, having lost all his near
and dear, undertakes the lonely trek to the heavens, with just a
faithful dog following him. At the gates of heaven, he is welcomed,
but is informed that the dog cannot come in. The just king then
rejects the invitation to heaven, if the faithful companion has to
be abandoned. All this again refers to the constellation Orion the
Hunter trailed by Sirius the dog, these being just outside the
heavenly zodiacal belt, the path of the
Sun, or the abode of Vishnu, the Sun.\\
One of the oldest of the Indian epics is the Ramayana which is built
around the story of the king Rama, an embodiment of all that is
moral and just. Rama or more fully Ramachandra, Chandra meaning the
Moon is wedded to Sita, again a symbol of perfection and daughter of
the king Janaka, literally king of man. Janaka finds Sita in the
fields while ploughing. If we sift through the symbolism and
imagery, this relates to the first agriculture practised by human
beings who thereby relinqued the hunter gatherer lifestyle. Sita the
produce of the Earth that is the harvest is wedded to the Moon,
which symbolises the months and seasons. This was the beginning of
the earliest calendar which was based on lunar months. The invention
of the calendar in turn, was forced on man, necessitated by
agricultural activity. It must be mentioned that, chronologically, there is
agreement with modern estimates of the start of agricultural activity after,
what scholars term, the e;pi-paleolithic age \cite{ck}.\\
The Ramayana is a description of events that took place at a time
when humans coexisted with a sub human species called Vanaras,
literally not fully human, but in myth considered to be monkeys. The
Vanaras used stone implements and clubs unlike the humans who used
the bow and arrow. This would sound strange to us today. But the
archaeological excavations near Nevali Cori \cite{hauptman}, show
exactly this coexistence. Stone age, more precisely neolithic
implements were found along with megalithic structures, that is
carved pillars, well planned halls and habitations and intricate
sculptures. The archaeological date of this civilization is around
7500 B.C., a date that coincides with the author's astronomical
chronology of portions of the Rig Veda and the Ramayana \cite{ck}.
We must remember that cultures generally coexist for a while with
succeeding cultures, before dying out. For example even today, there
are bushmen in Africa, the Amazonian tribesmen in Brazil and the
aborigines in Indian, Andaman and Nicober Islands.\\
In contrast the epic Mahabharata describes events which clearly took
place much later, in the iron age.\\
In the context of the Ramayana we would like to point out two other
interesting facts: The first relates to the ancient Indian
mythological concepts of the ten manifestations or avatars. This
actually corresponds quite closely to the evolutionary pattern of
life on the Earth. Thus the first manifestation or avatar was that
of the fish symbolizing the origin of life in the sea. The second
manifestation was that of the tortoise, symbolizing the amphibious
character of the next stage of evolution. The third manifestation
was that of a boar symbolizing the appearance of mammals. Then we
move through hairy man like beings and pygmies. Coming to the
Ramayana, the manifestation of the human Parusurama is mentioned,
whose weapon was an axe. Parasu stands for an axe and is the origin
of the Persues cluster in the sky. Parasurama, the axe wielder lived
up to the time of Rama, the next manifestation in which bows and
arrows were used. According to mythology Parasurama was overcome by
Rama. This is symbolic of the fact that the stone axe of the stone
age gave way to the more modern bow and arrow. The manifestation of
Rama was followed by Krishna one of the heroes of the epic
Mahabharata. Krishna belonged to the early prehistoric times and the
iron age. In fact he used several of these iron weapons. Moreover,
the fact that Rama and Krishna belonged to different, though succeeding
eras, is explicitly mentioned.\\
Another correlation in connection with the Ramayana is the bridge
that was built by Rama and his army of Vanaras, taken to be monkeys,
between India and Srilanka. It came as a complete surprise in the
last couple of centuries and more particularly a few decades ago,
particularly with satellite imaging techniques, that indeed there is
such a bridge like formation connecting India to Srilanka at exactly
the same place, as described in the epic. The probability of this
being an accident would be the same as that of several stones thrown
in random directions to hit pre assigned spots on the ground, that
is zero. It is easy to see that the beautiful metaphor of a bridge
being built was woven around the knowledge
of the existing formation.\\
Indeed a close study of some of the ancient literature of India
reveals a surprisingly accurate knowledge, not only of the skies but
also of geography. For example the Markandeya Purana \cite{pargitar}
describes on the one hand, the Earth as being flattened near the
poles, and on the other hand goes on to describe various lands
including Samar Kanda the modern Samarkand and several rivers
including some nearly forgotten rivers in India like the Vamsadhara
and Nagavalli.\\
Many mythological depictions in fact describe the drama in the
heavens as guessed by Mukherjee in the nineteenth century
\cite{kalinath}. An interesting example would be that of the Hindu
Goddess Saraswati who rides a swan and plays a stringed musical
instrument. The celestial Saraswati is the Milky Way flowing over
the Swan Cygnus with the instrument Lyra the Harp. This kind of
picturization is quite common. For example the Hindu Goddess Durga
who appears as a virgin, riding a lion, can be immediately
identified with the
constellation Virgo the virgin atop Leo the lion and so on \cite{ck}.\\
An interesting mythological story is that of the ancient Indian king
Bhageerath who saw the spirits of many of his ancestors in a
grieving state. These ancestors were all children of the king
Sagara, literally sea and they requested Bhageerath to bring to them
the waters of the celestial Ganga or Ganges, for their redemption.
This is a beautiful tale which describes the many travails and
tribulations that Bhageerath had to undergo to carry out the wishes
of his ancestors. First he had to convince the celestial Ganges to
come down to the Earth. But then the Earth would break up with the
impact. So he had to plead and persuade the God Siva to take the
impact on to himself. Siva agreed but soon thereafter got annoyed
and tied up the falling waters in his hair knot or sikhara,
literally peak. Then Bhageerath had to once again plead and persuade
Siva to release these waters which he did. After many more travails,
finally Bhageerath could bring the water to his ancestors, the
children of Sagara the sea at the Bay of Bengal. To this day the
event is celebrated there annually. Shorn of the mythology and
imagery, this is the tale of the water which falls down on the
Himalayas and gets frozen as snow on those lofty peaks, one of which
is called the Sivalik Mountain range to this day. The snows then
melt and the waters pour down and flow as the river Ganga and its
many tributaries - in fact the river feeds the drying up
tributaries, Bhageerath's ancestors, the children of the sea.
Finally the Ganga ends up in the Bay of Bengal.\\
Returning to the Mahabharata, it is mentioned there that the warrior
hero Arjuna accompanied by his companion Krishna encounters a being
termed an "Asura", and named Maya. They befriend Maya, who in return
for the friendship, builds a castle for Arjuna. According to Indian
mythology, the Asuras has the planet Venus (Sukra) as their
preceptor while Arjuna and his folk had the planet Jupiter
(Brihaspati) as their preceptor. According to Mayan (or Olmec)
legend, they were visited by people who were both fair and dark -
Arjuna literally means fair and Krishna literally means dark. (These
names were in fact, epithets). Further these legends suggest that
the fair warrior was ambidextrous - which in fact Arjuna was known
to be. Final insight into the meaning of all this comes from two
separate facts. According to the Panchasiddhantika of Varahamihira,
C.500 A.D., a compilation of five calenderic astronomical traditions
then known \cite{pancha}, the Asuras were antipodal people - indeed
India and Mexico are on opposite sides of the meridian. Moreover it
is known that the ancient Mexicans were obsessed with Venus and they
had detailed observations of this planet, preserved today in the
Dresden Codex. On the other hand, to this day, the cycle of years in
India is the Jovian cycle of sixty years of the planet Jupiter.
Moreover both these cultures belong to the same early prehistoric
period. Thus the supposedly mythological Mahabharata legend
describes the encounter of these two cultures. All this would explain the
puzzling presence of some Hindu motifs, e.g., the elephant, or tortoise
carrying twelve pillars, in Mexican symbolism.\\
A few comments are called for. As mentioned, the composers of these
myths had a surprising knowledge of physical events, including for
example the water cycle. The Sun is referred to in the Aditya
Hrudayam or Hymn of the Sun (a part of Ramayana) as one who drinks
the water on the Earth and then pours it down back on the Earth.
There are innumerable such insights strewn across this vast ocean of
literature. Secondly some scholars concluded, without any evidence,
that even such a simple concept as the zodiacal constellations and
other constellations were of Greco Babylonian origin. By a detailed
analysis the author has shown that indeed they were known long
before this period, and many of these concepts were in fact
inherited from an earlier legacy \cite{ck}.

\end{document}